\documentclass[a4paper,11pt]{article}
\usepackage{pos}
\usepackage{macros}
\usepackage{booktabs}
\usepackage[utf8]{inputenc}	
\usepackage[T1]{fontenc}		
\usepackage[english]{babel}		
\usepackage[rflt]{floatflt}
\newcommand{\halfwidth}{0.48\linewidth}
\newcommand{\fullwidth}{0.75\linewidth}
\usepackage[tableposition=top]{caption}

\usepackage[firstpageonly=true,angle=0,hpos=18.2cm,vpos=5.8cm,text=MS-TP-22-54,vanchor=t,hanchor=l,fontsize=12pt, color=black]{draftwatermark}

\title{On improvement of the axial-vector current with stabilised Wilson fermions}

\author[a]{P. Fritzsch}
\author[b]{J. Heitger}
\author*[b]{J. T. Kuhlmann}

\affiliation[a]{School of Mathematics, Trinity College, College Green, Dublin, Ireland}
\affiliation[b]{Institut für Theoretische Physik, Westfälische Wilhelms-Universität Münster,\\
	Wilhelm-Klemm-Straße 9, 48149 Münster, Germany}

\emailAdd{fritzscp@tcd.ie}
\emailAdd{heitger@uni-muenster.de}
\emailAdd{j.kuhlmann@uni-muenster.de}

\abstract{We report on the determination of the improvement coefficient $\ca$ for the non-singlet axial-vector current $A^{a}_{\mu}(x)$ in the framework of stabilised Wilson-Clover fermions within three flavour lattice QCD.
	This is done by requiring the PCAC relation to hold for two different pseudo-scalar states. To generate these states, wavefunctions altering spatial structures on the boundaries of a Schrödinger functional lattice are employed and some variations of the (previously applied) wavefunction method are explored.
	The improvement coefficient is determined on a few ensembles for a range of gauge couplings that is potentially useful for future applications.
	Preliminary results on the renormalisation constants $\zv$ for the vector current and $\za$ for the axial-vector current are also presented.}

\FullConference{%
  The 39th International Symposium on Lattice Field Theory (Lattice2022),\\
  8-13 August, 2022 \\
  Bonn, Germany 
}

\begin{document}
\maketitle
\section{Introduction}
Leptonic pseudo-scalar hadronic decays have a significant impact on precisely estimated entries of the CKM matrix and therefore are of great interest for indirect tests of the Standard Model. Since these decays involve the QCD hadronic matrix element of the axial-vector current
\begin{equation}
	\bra{0}A^a_\mu(x)\ket{\pi^b(p)}=-ip_\mu f_\pi\,\delta^{ab}e^{-ipx}\,,
\end{equation}
lattice QCD is the favoured way to reliably estimate the low-energy part of the theory contribution to these tests.

As is well-known, Wilson fermions, while having many advantages, break chiral symmetry and need to be improved to remove its leading $\rmO(a)$ discretisation errors. For this purpose, resorting to the well-established Symanzik improvement programme \cite{Luescher1996a}, the fermion action receives an extra term known as the Sheikholeslami-Wohlert term. Quark bilinears also need to be improved by adding suitable extra terms. Within this programme it turns out that the improvement term of the axial-vector current is proportional to the derivative of the pseudo-scalar density. In a mass independent scheme, the (multiplicatively) renormalised and improved current for degenerate quarks is of the form
\begin{equation}
	(\arimpr)^a_\mu(x)=\za(1+\ba m_q)(\aimpr)^a_\mu(x)
	=\za(1+\ba m_q)\left[A^a_\mu(x)+\ca\partial_\mu P^a(x)\right]
	\label{eq:axial_curr_RI}
\end{equation}
with the axial-vector current and pseudo-scalar density being defined as
\begin{equation}
	A^a_\mu\left(x\right) =
	\bar{\psi}\left(x\right)\gamma_{\mu}\gamma_5\frac{\lambda^a}{2}\psi\left(x\right)
	\text{~~~and~~~}
	P^a =
	\bar{\psi}\left(x\right)\gamma_5\frac{\lambda^a}{2}\psi\left(x\right)\,.
	\label{eq:currents}
\end{equation}
In practice, for a non-perturbative determination of $\ca$ and $\za$ one works in the chiral limit of vanishing quark masses where the $\ba$-term vanishes.

Here we report on our efforts to improve and renormalise the axial-vector current for three-flavour lattice QCD with stabilised Wilson fermions (SWF) \cite{Francis2019}.
For this computation we employ ensembles with fairly large volumes ($L\approx 3\,\rm{fm}$), Schrödinger functional boundary conditions and lattice constants in a range between $a = 0.12\,\rm{fm}$ ($\beta = 3.685$) and $a = 0.055\,\rm{fm}$ ($\beta = 4.10$) in the chiral limit as well as at the symmetric point of fully degenerate sea and valence quark masses.
Since SWF provide a new $\rmO(a)$ discretisation of QCD, their improvement coefficients and renormalisation constants differ from the known $\nf = 3$ values (see e.g. \cite{Bulava2013,Bulava2015,Bulava2016,DallaBrida2019}) for ordinary Wilson-Clover fermions with the same, tree-level Symanzik-improved gauge action.
\section{Stabilised Wilson formulation}
Numerical simulations with the Wilson-Clover Dirac operator
\begin{equation}
	D_{{\rm W}} = \frac{1}{2}\left\lbrace\gamma_{\mu}\left(\nabla_{\mu}^*+\nabla_{\mu}\right)-\nabla_{\mu}^*\nabla_{\mu}\right\rbrace +\csw\frac{i}{4}\sigma_{\mu \nu}F_{\mu\nu}+m_0
\end{equation}
require stabilising measures as proposed in \cite{Francis2019}. Based on the observation that the clover term in $D_{{\rm W}}$ hinders stable inversion of the Dirac matrix, an alternative form of the Dirac operator was introduced, 
in which the on-site terms are exponentiated, i.e.
\begin{equation}
	M_0+\csw\frac{i}{4}\sigma_{\mu \nu}F_{\mu \nu}\quad\rightarrow \quad M_0\exp\left(\frac{\csw}{M_0}\frac{i}{4}\sigma_{\mu \nu}F_{\mu \nu}\right)\,,
\end{equation}
with $M_0 = m_0+4$.
The exponential ensures an intrinsic bound from below in this term. As a consequence, the Dirac operator does receive fewer near-zero eigenvalues, stabilising the calculation. This approach was originally developed for master-field simulations \cite{Francis2019}, but it turns out to also be beneficial for light quarks, coarse gauge fields and large lattices \cite{Cuteri2022}. This is the situation we are focussing on.

While the Sheikholeslami-Wohlert coefficient $\csw$ for the new action has already been determined for a range of couplings \cite{Francis2019,Cuteri2022}, the improvement coefficients for quark-bilinears such as the axial-vector current are yet to be computed.
\section{Improvement of the axial-vector current}
Along the lines of \cite{Morte2005, Bulava2015}, using the Schrödinger functional framework, one can derive an expression for the coefficient $\ca$. To do so, we introduce the correlation functions
\begin{equation}
	\fa(x_0;\omega) = -\frac{a^3}{3L^6} \sum_{\vec{x}} \langle A^a_0(x) O^a(\omega)\rangle
	\text{\quad and\quad}
	\fp(x_0;\omega) = -\frac{a^3}{3L^6} \sum_{\vec{x}} \langle P^a(x) O^a(\omega)\rangle
	\label{eq:f_A_and_f_P}
\end{equation}
with the boundary source operator at $x_0 = 0$ being defined as
\begin{equation}
	O^e(\omega)=a^6\sum_{\vec{y},\vec{z}}\bar{\zeta}(\vec{y})\gamma_5\frac{\lambda^e}{2}\omega(\vec{y}-\vec{z})\zeta(\vec{z})\,,
	\label{def:bound_PS_source_discr}
\end{equation}
where $\zeta$ is the boundary quark field. Similar correlation functions can also be introduced with sources at $x_0 = T$. These are built from boundary operators $O'^f(\omega)$ and complementary boundary quark fields $\zeta '$.
In this context, $\omega$ denotes an arbitrary spatial structure on the boundary that we will call wavefunction. The first step in our analysis is to find suitable wavefunctions such that the ground and first excited state can be readily extracted from the correlators. These are approximated as in \cite{Morte2005,Bulava2015} by constructing linear combinations of basis wavefunctions. Here we consider a set of five functions of the form
\begin{equation}
	\omega^{\text{b}1} = \mathrm{e}^{-\frac{r}{a_0}}\,, ~\omega^{\text{b}2} = r~ \mathrm{e}^{-\frac{r}{a_0}}\,, ~\omega^{\text{b}3} = \mathrm{e}^{-\frac{r}{2a_0}}\,, ~\omega^{\text{b}4} = \text{const}\,,
	~\omega^{\text{b}5} = r^2~\mathrm{e}^{-\frac{r}{a_0}}\,,
\end{equation}
which resemble those of the hydrogen atom, with $r = |\vec{x}-\vec{y}|$ and $a_0$ parametrising the spatial extent of the function. To find a well-suited linear combination, the boundary-to-boundary correlator,
\begin{equation}
	\Fone(\omega^{{\rm b}i},\omega^{{\rm b}\!j}) = -\frac{1}{3L^6}\langle O'^a(\omega^{{\rm b}\!j}) O^a(\omega^{{\rm b}i})\rangle\,,
	\label{eq:F_1}
\end{equation}
which for our purposes can be viewed as a $5\!\times\!5$-matrix in wavefunction space, is evaluated. We can pick any subset of indices as a submatrix\footnote{These need to be at least $3\!\times\!3$-submatrices for the eigenvectors to be fully independent of each other.} and calculate its  eigenvectors, sorted by eigenvalue. The \glqq standard\grqq~ choice that was previously used in refs. \cite{Morte2005,Bulava2015} can be regained by setting $i,j\in\{1,2,3\}$. In our setup it turned out that with this choice $\ca$ inherits some ambiguity that may be avoided by other linear combinations. Hence we also explored the submatrix with $i,j\in\{1,2,4\}$.
Exploiting the overlap information of the basis wavefunctions to approximately prepare the ground and first excited state then amounts to projecting the correlators $\fa$ and $\fp$ with given basis wavefunctions onto the eigenvectors belonging to the largest two eigenvalues.

The improvement coefficient $\ca$ is determined in the following way: The starting point is the PCAC relation 
\begin{equation}
	\partial_{\mu} \langle A^a_\mu(x) O^a \rangle = 2\mpcac~\langle  P^a(x)O^a\rangle\,,
	\label{eq:PCAC_relation}
\end{equation}
which follows from an invariance under chiral rotation and holds as an operator identity in the continuum theory. Therefore, all effects that break it on the lattice must stem from the discretisation. Requiring the identity to be satisfied up to $\rmO(a^2)$ cut-off effects amounts to inserting the improved current from eq. \eqref{eq:axial_curr_RI} into this relation and thus yields for the $\rmO(a)$ improved PCAC quark mass
\begin{equation}
	a \mpcac \equiv \frac{a~\partial_{\mu}\langle  (\aimpr)^a_\mu(x)O^a\rangle}{2~\langle P^a(x)O^a\rangle}+\rmO(a^2)\,,
\end{equation}
provided $\ca$ is fixed non-perturbatively. As detailed in the previous works, it can be rewritten in terms of the Schrödinger functional correlation functions from above as
\begin{equation}
	a~\mpcac(x_0;\omega^i) = \frac{a~\hat{\partial}_{0} \fa(x_0;\omega^i)+\ca~a^2~\hat{\partial}^2_{0} \fp(x_0;\omega^i)}{2~ \fp(x_0;\omega^i)}\,.
\end{equation}
Decomposing the PCAC mass as
\begin{equation}
	\mpcac(x_0;\omega^i) = r(x_0;\omega^i)+\ca~a~s(x_0;\omega^i)\,, 
\end{equation}
with
\begin{equation}
	r(x_0;\omega^i) = \frac{\hat{\partial}_{0} \fa(x_0;\omega^i)}{2~ \fp(x_0;\omega^i)}
	\text{\quad and\quad}
	s(x_0;\omega^i) = \frac{\hat{\partial}^2_{0} \fp(x_0;\omega^i)}{2~ \fp(x_0;\omega^i)}\,,
	\label{eq:r_and_s}
\end{equation}
and assuming that it holds for the two lowest states in the pseudo-scalar channel (isolated as described above) labelled by indices $i = 0,1$, we can solve for $\ca$:
\begin{equation}
	\ca(x_0)= -\frac{r(x_0;\omega^{1}) - r(x_0;\omega^{0})}{a~(s(x_0;\omega^{1})-s(x_0;\omega^{0}))}=-\frac{\Delta r(x_0)}{a~\Delta s(x_0)}\,.
	\label{def:c_A_condition}
\end{equation}
Note that, as observed in \cite{Morte2005, Bulava2015}, the precision of the estimate for $\ca$ is dominated by the statistical error of $a\Delta s$. Therefore an examination of the $x_0$-dependence of the functions $\Delta r$ and $a\Delta s$ may be helpful.
As one expects $\ca(x_0)$ in eq. \eqref{def:c_A_condition} to develop a plateau over a certain range of $x_0$, a suitable plateau range needs to be identified, in which the noise in $\Delta r$ and $a\Delta s$ is small. 
\section{Renormalisation}
For the calculation of the renormalisation factor $\za$ we adopt the strategy explained in \cite{Morte2005a,Bulava2016}. We only recall here that the calculation involves a Ward identity originating from an infinitesimal chiral rotation of the fields entering the expectation value
\begin{equation}
	\langle A_{\nu}^b(y)O_{\text{ext}}\rangle=\int \mathcal{D}[\bar{\psi},\psi,U] A^b_{\nu}(y)O_{\text{ext}}e^{-S}\,,
\end{equation}
where we include the current into the chiral rotation and obtain from the variational principle
\begin{equation}
	0 = \langle\delta A^b_{\nu}(y)O_{\text{ext}}\rangle-\langle\delta S A^b_{\nu}(y)O_{\text{ext}}\rangle\,.
	\label{eq:van_var_exp_value}
\end{equation}
For $O_{\text{ext}}$ which we set
\begin{equation}
	O_{\text{ext}} = O^{ef}_{\text{ext}} =O^e (O')^f\,,
\end{equation}
where $O^e$ and $(O')^f$ are the pseudo-scalar boundary sources as defined in eq. \eqref{def:bound_PS_source_discr}.

\begin{figure}
	\centering
	\includegraphics[width=\fullwidth]{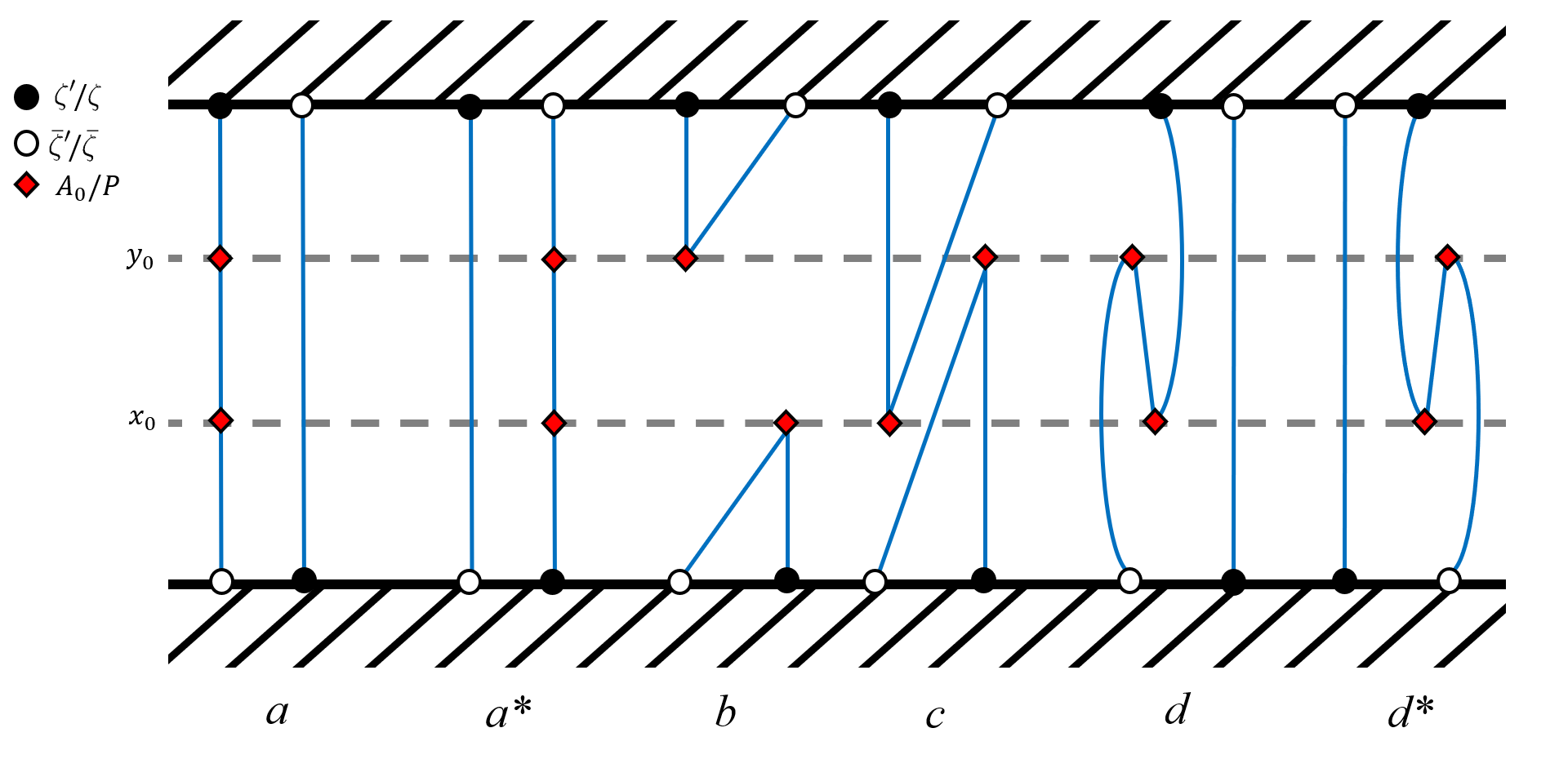}
	\vspace{-0.3cm}
	\caption{The six non-vanishing Wick contractions contributing to the 4-point functions $F_{{\rm XY}}$.}
	\label{fig:4p-contribs}
\end{figure}

Omitting all intermediate steps, translating the expressions into their lattice counterparts and formulating them in terms of renormalised, improved Schrödinger functional correlators, one can finally isolate $\za$ as
\begin{equation}
	\za = \sqrt{\frac{\Fone}{\faaimpr(x_0,y_0)-2\mpcac \tilde{F}_{\rm{PA}}^{\rm I}(x_0,y_0)}}\,,
	\label{eq:Z_A_full}
\end{equation}
where $\faaimpr$ is defined as 
\begin{equation}
	\faaimpr=\faa(x_0,y_0)
	+a \ca \left(\partial_{x_0} \fpa(x_0,y_0)+\partial_{y_0} \fap(x_0,y_0)\right)
	+a^2 \ca^2 \left(\partial_{x_0}\partial_{y_0} \fpp(x_0,y_0)\right)
\end{equation}
and $\tilde{F}_{\rm{PA}}^{\rm{I}}$ as 
\begin{equation}
	\tilde{F}^{\rm I}_{\rm{PA}}(x_0,y_0) = \tilde{F}_{\rm{PA}}(x_0,y_0)+\ca\partial_{y_0} \tilde{F}_{\rm{PP}}(x_0,y_0)\,,
\end{equation}
with
\begin{equation}
	F_{\rm{XY}}(x_0,y_0) \equiv -2a^6\sum_{\vec{x}, \vec{y}} \langle X^1(x) Y^2(y)  O^2 (O')^1 \rangle
\end{equation}
and
\begin{equation}
	\tilde{F}_{\rm{XY}}(x_0,y_0) \equiv a\sum_{z_0 = x_0}^{y_0}w(z_0)F_{\rm{XY}}(z_0,y_0)\text{\quad with\quad}	w(z_0)=
	\begin{cases}
		\frac{1}{2}\;\text{if~} z_0 = x_0\text{\;or\;}z_0 = y_0\\
		1\;\text{else}
	\end{cases}\,.
\end{equation}
Each of the 4-point functions receives contributions from six different diagrams shown in fig. \ref{fig:4p-contribs}. 
Two of these are disconnected and can be evaluated as a product of two separate 2-point functions.

Additionally, in this calculation we consider the vector current which enters the following Ward identity:
\begin{equation}
	\int\mathrm{d}^3y\langle V^c_0(y)O^e (O')^f\rangle = i f^{ce}_g \langle O^g (O')^f\rangle \,,
	\label{eq:vec_ward_id}
\end{equation}
where $O^e$ and $(O')^f$ are defined as before.
From this, a simple expression for $\zv$ can be derived~\cite{Luescher1996c}:
\begin{equation}
	\zv = \frac{\Fone}{\fv}+\rmO(a^2)\,,
	\label{eq:Z_V}
\end{equation}
in which $\fv$ is 
\begin{equation}
	\fv (x_0) = i\sum_x \varepsilon_{abc}\langle O'^a V^b_0(x_0) O^c \rangle \,.
\end{equation}
\section{Ensembles and error analysis}
After having discussed the theoretical aspects, we introduce the ensembles used in the numerical computations so far. To make our results on $\ca$, $\za$ and $\zv$ useful for other groups, the range of ensembles covered are at the same lattice spacings $a$ as those generated by OpenLat\footnote{see also: \url{https://openlat1.gitlab.io/}} \cite{Cuteri2022}, only differing in boundary conditions.
The ensembles used here are listed in tab. \ref{tab:ensembles}.
Our error analysis is done employing the $\Gamma$-method \cite{Wolff2003} using the python-implementation \texttt{pyerrors} described in \cite{Joswig2022}.
\begin{table}
	\caption{Ensembles with Schrödinger functional boundary conditions used in this project. The mass point indicates to which line of constant physics the ensemble belongs.}
	\label{tab:ensembles}
	\vspace{-0.5cm}
	\begin{center}
		\begin{tabular}{ccccccccccccccc}
			\toprule
			&$T$&&$L^3$&&$\beta$&&$\kappa_{u/d/s}$&&$a$ [fm]&&mass point&&\# of configs&\\
			\midrule
			&$24$&&$16^3$&&$3.80$&&$0.1392500$&&$0.095$&&chir.&&$2875$&\\
			&$24$&&$24^3$&&$3.80$&&$0.1392500$&&$0.095$&&chir.&&$2523$&\\
			&$32$&&$32^3$&&$3.80$&&$0.1392500$&&$0.095$&&chir.&&$1167$&\\

			&$24$&&$24^3$&&$3.685$&&$0.1394400$&&$0.120$&&symm.&&$540$&\\

			&$32$&&$32^3$&&$3.80$&&$0.1389630$&&$0.095$&&symm.&&$956$&\\

			&$56$&&$56^3$&&$4.10$&&$0.1380000$&&$0.055$&&symm.&&$33$&\\
			\bottomrule
		\end{tabular}
	\end{center}
\end{table}
As stated in the introduction, we are interested in two lines of constant physics: at the chiral point of (almost) vanishing quark masses and at the symmetric point of fully degenerate massive quark flavours.
\section{Results}
\begin{figure}
	\begin{tabular}{cc}
		\includegraphics[width = \halfwidth]{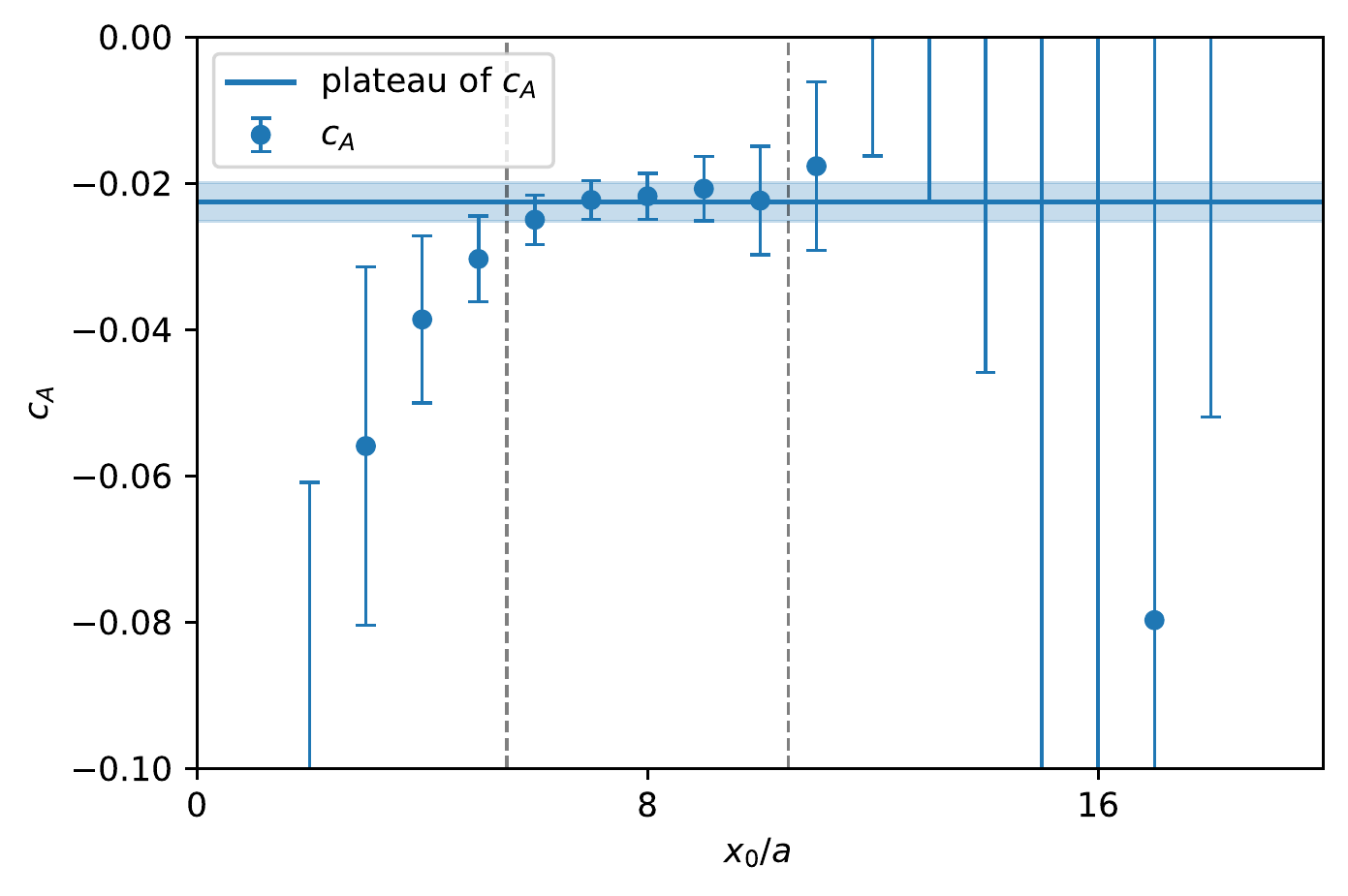}&
		\includegraphics[width = \halfwidth]{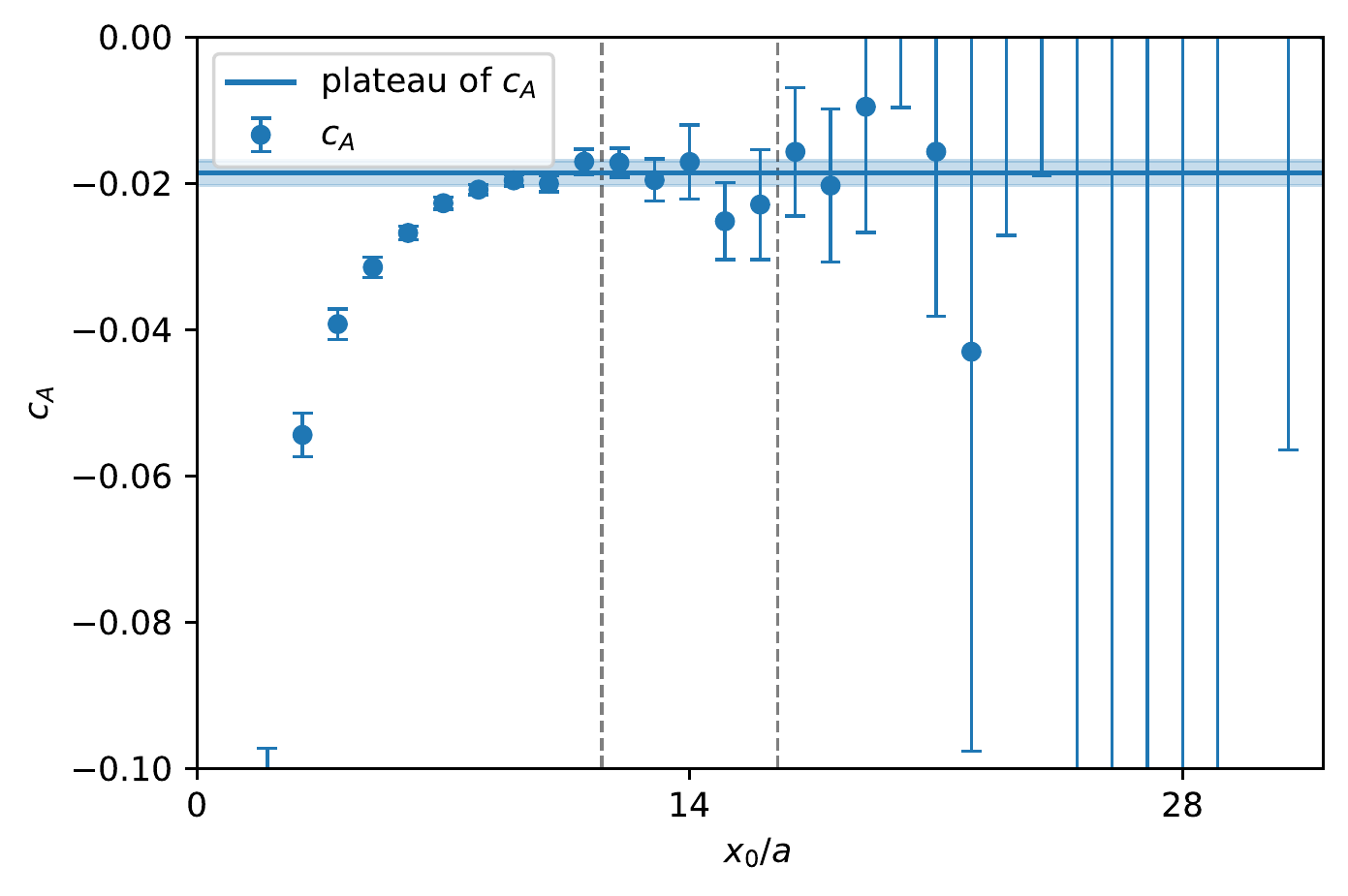}
	\end{tabular}
	\vspace{-0.3cm}
	\caption{Preliminary results for $\ca$ at $\beta = 3.80$ ($a \approx 0.094\,\text{fm}$, left) at the chiral point and $\beta = 4.10$ ($a \approx 0.055\,\text{fm}$, right) at the symmetric point with the standard projection.}
	\label{fig:c_A_T32_T56_std}
\end{figure}
\begin{floatingfigure}{0.5\linewidth}
	\centering
	\includegraphics[width = \halfwidth]{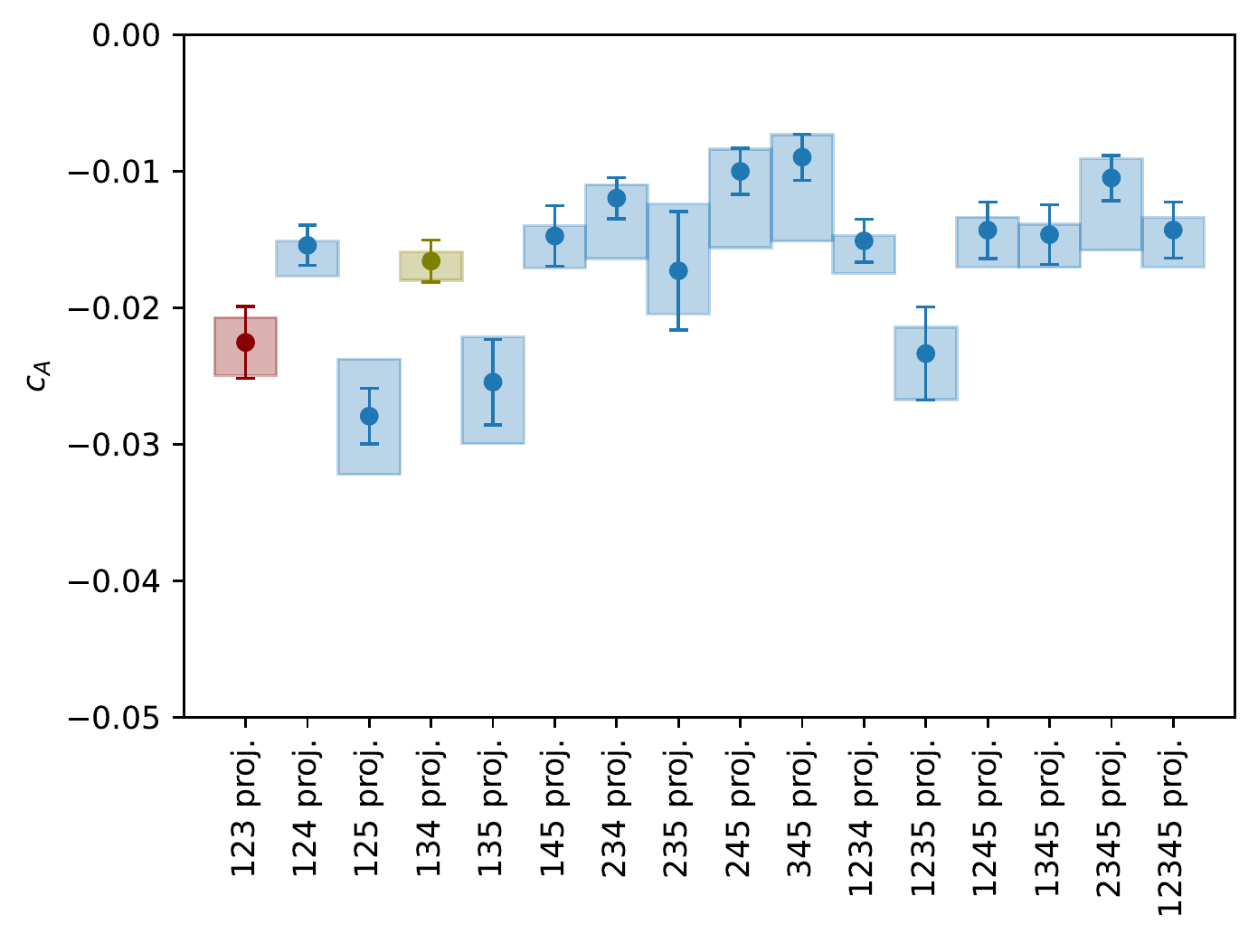}
	\vspace{-0.3cm}
	\caption{Example for the possible projections on the $32^4$ lattice at $\beta = 3.80$ at the chiral point. Data points shown here are the plateau-values of $x_0\in [T/4-2, T/4+2]$. The boxes indicate the spread of $\ca$ in the plateau region. The standard projection is marked in red, while the new, preferred projection is marked in green.
	}
	\label{fig:c_A_T32_chir_poss_proj}
\end{floatingfigure}

In the analysis for $\ca$ we have seen that the quality and longevity of the plateau may be problematic in the large-volume setup adopted here. For the preliminary results we set the plateau region as $x_0 \in [T/4-2, T/4+2]$, which works out fine in most cases, as can be seen for the two representative examples in fig. \ref{fig:c_A_T32_T56_std}.

As mentioned above, it has also been seen in our tests that the standard projection (used in \cite{Bulava2015,Morte2005}) is not always optimal. An alternative was found by scanning all possible projections which can be built after diagonalising $\Fone$ on various (subsets of) basis wavefunctions. The outcome of these scans can be inferred from fig. \ref{fig:c_A_T32_chir_poss_proj}. While the standard projection shown in red, assessed in isolation, yields a legitimate value, the majority of the other projections are placed systematically around a different value. This could hint at an overestimation of $\ca$ (in magnitude) when using the standard projection. However, the new, preferred projection exhibits smaller errors and yields results consistent with most other projections, thus appearing to be less affected by systematics due to the choice of the wavefunction basis. The spread of $\ca$ in the plateau region (shown in fig. \ref{fig:c_A_T32_chir_poss_proj} by the vertical extent of the boxes) is also smaller for this projection, which endorses the better quality of the associated plateau.

\begin{figure}
	\begin{tabular}{cc}
		\includegraphics[width = \halfwidth]{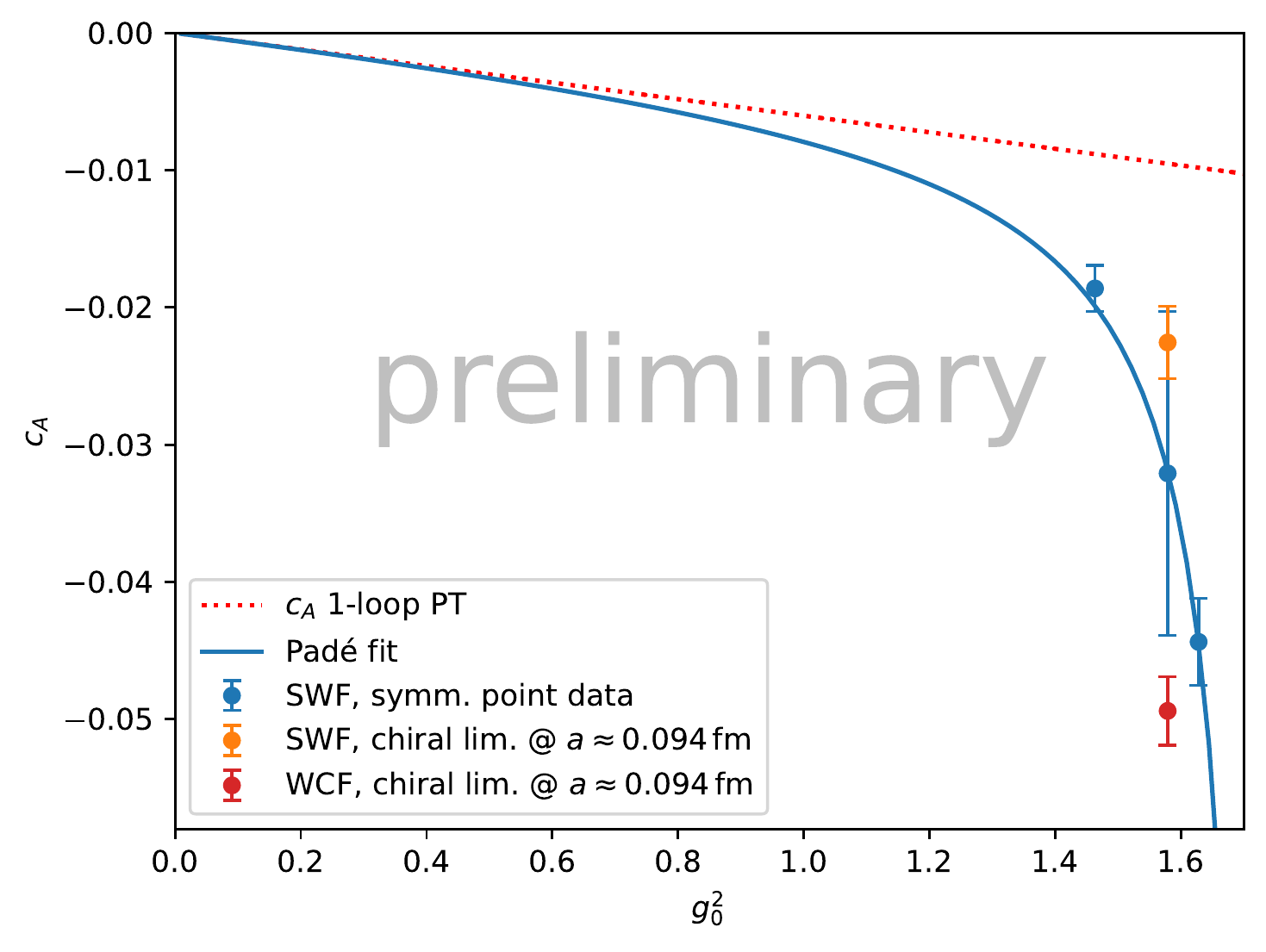}&
		\includegraphics[width = \halfwidth]{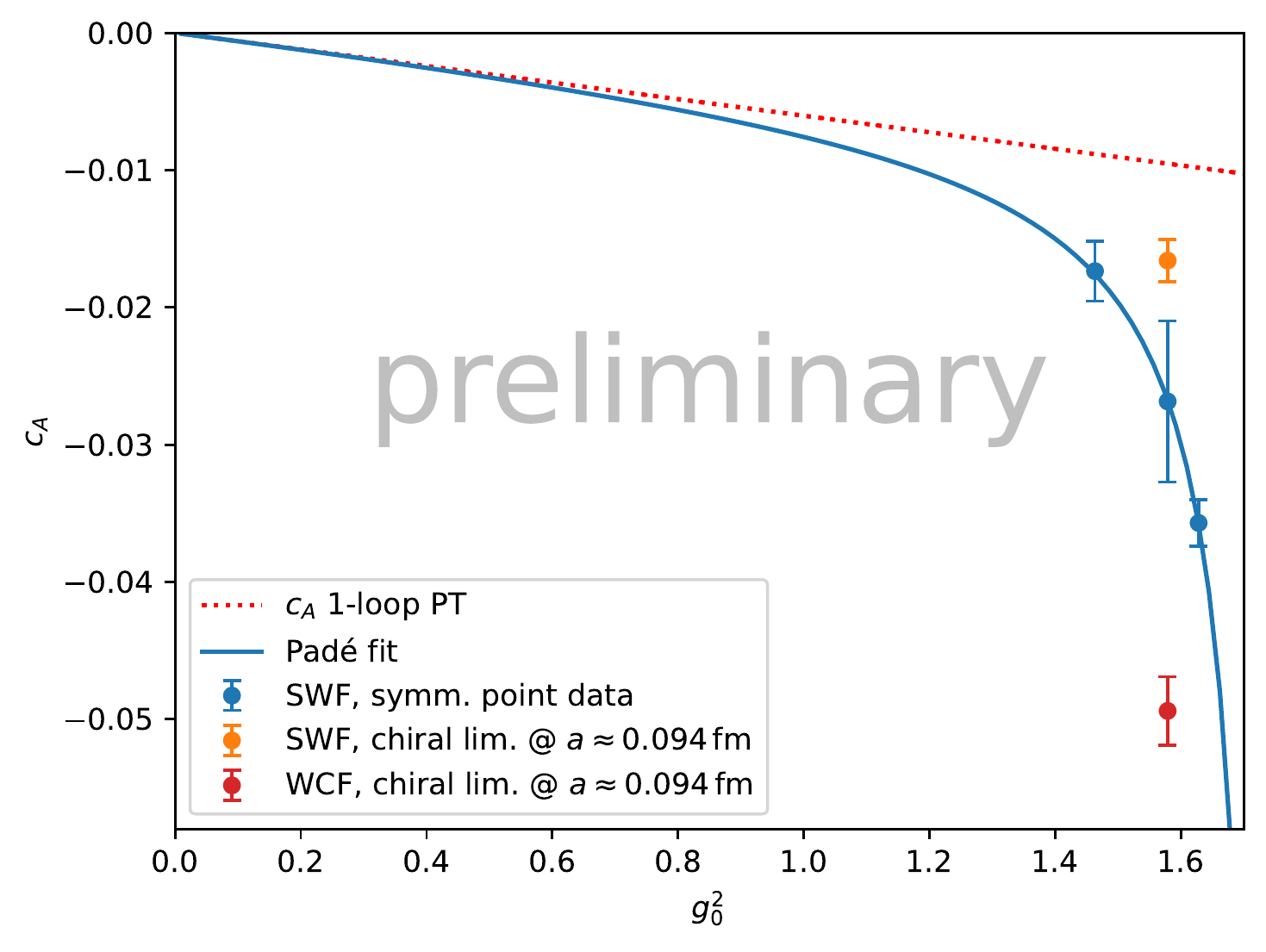}
	\end{tabular}
	\vspace{-0.3cm}
	\caption{Preliminary $g_0^2$-dependence of $\ca$ for ensembles at the symmetric point with $\beta = 3.685$, $\beta = 3.80$, $\beta = 4.10$ with the standard (left) and the preferred (right) projection. In orange, the point for the chiral case at $\beta = 3.80$ is shown. The red point gives $\nf=3$ result for traditional Wilson-Clover fermions from \cite{Bulava2015}.}
	\label{fig:beta_dep}
\end{figure}

In fig. \ref{fig:beta_dep} a preliminary interpolation in $g_0^2$ of the results on the symmetric point ensembles is displayed. So far, the ensembles studied do not yet allow for an analogous interpolation in the chiral case. However, we can see that in comparison with the results for ordinary Wilson-Clover fermions \cite{Bulava2015} the estimates for $\ca$ appear to be smaller in magnitude which may hint at smaller cut-off effects for stabilised Wilson fermions at the same lattice spacing. Of course, whether this is a general feature of this discretisation still needs to be confirmed in phenomenological applications.

For the renormalisation constants, preliminary results are listed in tables. \ref{tab:renorms_axial} and \ref{tab:renorms_vector}. These indicate that the estimates obtained are within the same ballpark as the ones for ordinary Wilson-Clover fermions in $\nf = 3$ \cite{Bulava2016,Heitger2020}. 
\begin{table}
	\caption{Preliminary estimates of renormalisation constants for the axial-vector channel. The labels f, c and m refer to full, connected and massive definitions of $\za$; see \cite{Bulava2016} for details.}
	\label{tab:renorms_axial}
	\centering
	\begin{tabular}{clccccc}
		\toprule
		&Ensemble&$\za^{{\rm f}}$&$\za^{\rm f, m}$&$\za^{{\rm c}}$&$\za^{\rm c, m}$&\\
		\midrule
		&$24\!\times\!16^3$, $\beta=3.80$, chir.&$0.7638(76)$&$0.7585(80)$&$0.753(20)$&$0.773(21)$&\\
		&$24\!\times\!24^3$, $\beta=3.80$, chir.&$0.779(15)$&$0.741(13)$&$0.7631(50)$&$0.7782(39)$&\\
		\bottomrule
	\end{tabular}
\end{table}
As the 4-point functions used here are prone to large statistical fluctuations, it seems more promising to determine $\za$ in the chirally rotated Schrödinger functional ($\chi \rm{SF}$) scheme \cite{DallaBrida2019,Sint2010}, because in this framework only 2-point functions are involved.

\begin{minipage}{0.42\linewidth}
		\includegraphics[width=\linewidth]{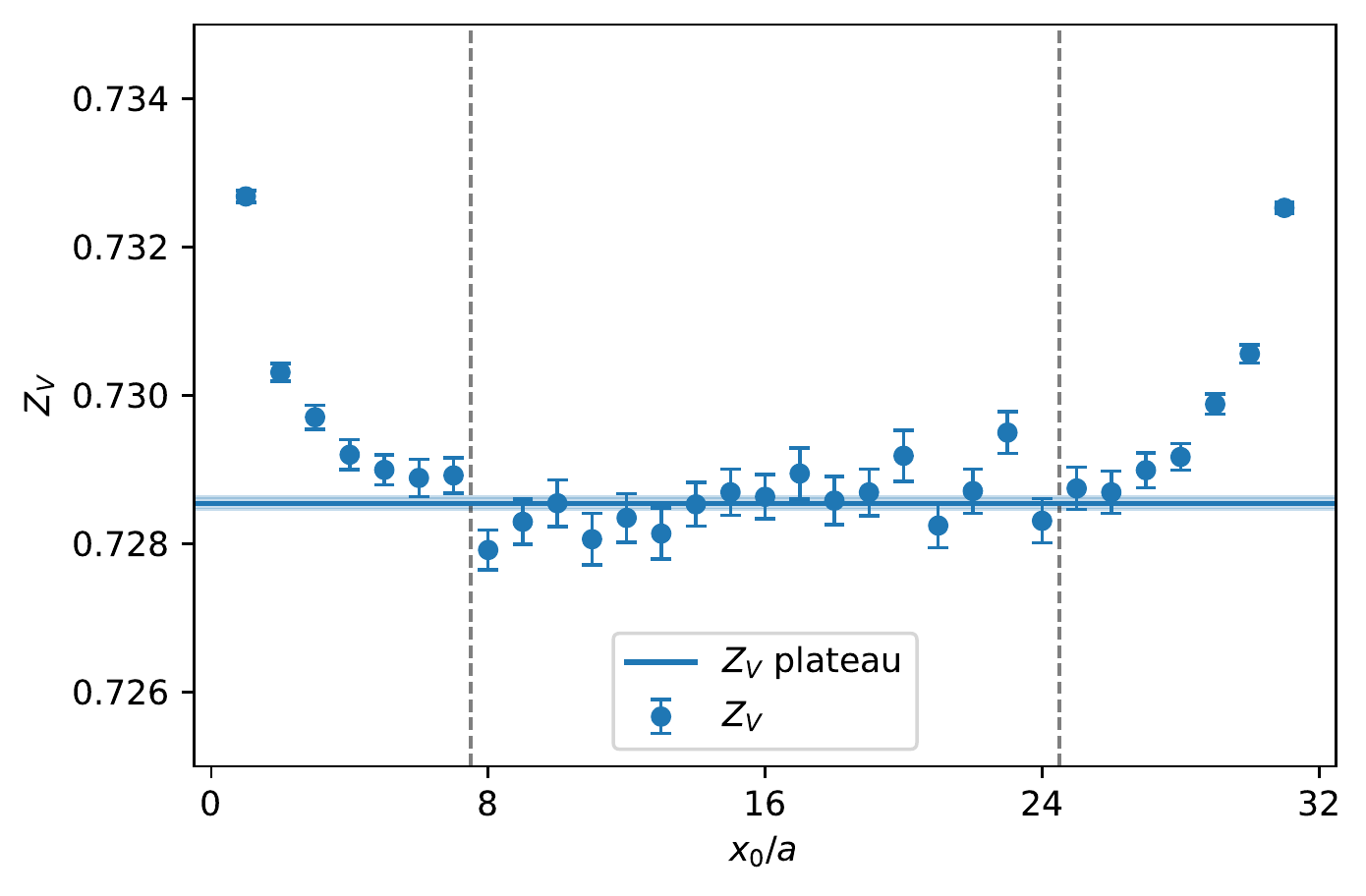}
		\vspace{-0.8cm}
		\captionof{figure}{Example for the calculation of $\zv$ on the chiral ensemble at $\beta = 3.80$ ($a\approx 0.094\,\text{fm}$).}
		\label{fig:Z_V_T32}
\end{minipage}
\hspace{0.5cm}
\begin{minipage}{0.42\linewidth}
	\captionof{table}{Preliminary estimates of renormalisation constants for the vector channel.}
	\label{tab:renorms_vector}
	\begin{tabular}{lc}
		\toprule
		Ensemble&$\zv$\\
		\midrule
		$24\!\times\!16^3$, $\beta=3.80$, chir.&$0.729141(94)$\\
		$24\!\times\!24^3$, $\beta=3.80$, chir.&$0.728716(56)$\\
		$32\!\times\!32^3$, $\beta=3.80$, chir.&$0.728546(81)$\\
		$24\!\times\!24^3$, $\beta=3.685$, symm.&$0.71209(13)$\\
		$32\!\times\!32^3$, $\beta=3.80$, symm.&$0.738044(53)$\\
		$56\!\times\!56^3$, $\beta=4.10$, symm.&$0.777900(17)$\\
		\bottomrule
\end{tabular}
\end{minipage}
\vspace{0.5cm}

The values for $\zv$ in tab. \ref{tab:renorms_vector} were extracted from plateau ranges like the one shown in fig. \ref{fig:Z_V_T32}. Due to the simple expression for $\zv$ in eq. \eqref{eq:Z_V}, which only consists of a ratio of two correlation functions and yields a good plateau quality in most cases, it can be determined to high precision. In the present, preliminary results, plateaus have been taken from $x_0 = T/4$ up to $x_0 = 3T/4$.
\vspace{0.3cm}
\section{Conclusions and outlook}
\vspace{-0.3cm}
We have seen that the Ward identity method in combination with Schrödinger functional boundary conditions can be readily implemented to determine $\ca$ for the stabilised Wilson-Clover fermions. As we work in larger volumes, some adaptions in the choice of the plateau range and wavefunctions turn out to be advantageous to reach the desired precision goal. To finally arrive at a proper interpolation over the whole range of interesting lattice spacings, measurements on further ensembles are still to be added.

When it comes to renormalisation, a precise extraction of $\zv$ for the vector current is straightforward, while the achievable accuracy for $\za$ remains to be seen. A promising route to enhance the final precision of $\za$ is to pursue the computational strategy within the $\chi$SF advocated and successfully applied in \cite{DallaBrida2019, Sint2010}.

\bibliographystyle{JHEP}
\bibliography{../../../Literatur/myLibrary.bib}

\providecommand{\href}[2]{#2}\begingroup\raggedright\begin{thebibliography}{10}

\bibitem{Luescher1996a}
M.~Lüscher, S.~Sint, R.~Sommer, and P.~Weisz, {\it {Chiral symmetry and O(a)
  improvement in lattice QCD}},  {\em Nucl. Phys. B} {\bf 478} (1996) 365,
  [\href{http://xxx.lanl.gov/abs/hep-lat/9605038}{{\tt hep-lat/9605038}}].

\bibitem{Francis2019}
A.~Francis, P.~Fritzsch, M.~Lüscher, and A.~Rago, {\it {Master-field
  simulations of O(a)-improved lattice QCD: Algorithms, stability and
  exactness}},  {\em Comput. Phys. Commun.} {\bf 255} (2019) 107355,
  [\href{http://xxx.lanl.gov/abs/1911.04533}{{\tt 1911.04533}}].

\bibitem{Bulava2013}
J.~Bulava and S.~Schaefer, {\it {Improvement of $N_f$ = 3 lattice QCD with
  Wilson fermions and tree-level improved gauge action}},  {\em Nucl. Phys. B}
  {\bf 874} (2013) 188, [\href{http://xxx.lanl.gov/abs/1304.7093}{{\tt
  1304.7093}}].

\bibitem{Bulava2015}
J.~Bulava, M.~D. Morte, J.~Heitger, and C.~Wittemeier, {\it {Non-perturbative
  improvement of the axial current in $N_f=3$ lattice QCD with Wilson fermions
  and tree-level improved gauge action}},  {\em Nucl. Phys. B} {\bf 896} (2015)
  555, [\href{http://xxx.lanl.gov/abs/1502.04999}{{\tt 1502.04999}}].

\bibitem{Bulava2016}
J.~Bulava, M.~D. Morte, J.~Heitger, and C.~Wittemeier, {\it {Non-perturbative
  renormalization of the axial current in $N_f = 3$ lattice QCD with Wilson
  fermions and tree-level improved gauge action}},  {\em Phys. Rev. D} {\bf 93}
  (2016) 114513, [\href{http://xxx.lanl.gov/abs/1604.05827}{{\tt 1604.05827}}].

\bibitem{DallaBrida2019}
M.~Dalla~Brida, T.~Korzec, S.~Sint, and P.~Vilaseca, {\it {High precision
  renormalization of the flavour non-singlet Noether currents in lattice QCD
  with Wilson quarks}},  {\em Eur. Phys. J. C} {\bf 79} (2019) 23,
  [\href{http://xxx.lanl.gov/abs/1808.09236}{{\tt 1808.09236}}].

\bibitem{Cuteri2022}
A.~S. Francis, F.~Cuteri, P.~Fritzsch, G.~Pederiva, A.~Rago, A.~Schindler,
  A.~Walker-Loud, and S.~Zafeiropoulos, {\it {Properties, ensembles and hadron
  spectra with Stabilised Wilson Fermions}},  {\em PoS} {\bf LATTICE2021}
  (2022) 118, [\href{http://xxx.lanl.gov/abs/2201.03874}{{\tt 2201.03874}}].

\bibitem{Morte2005}
M.~Della~Morte, R.~Hoffmann, and R.~Sommer, {\it {Non-perturbative improvement
  of the axial current for dynamical Wilson fermions}},  {\em JHEP} {\bf 03}
  (2005) 029, [\href{http://xxx.lanl.gov/abs/hep-lat/0503003}{{\tt
  hep-lat/0503003}}].

\bibitem{Morte2005a}
M.~Della~Morte, R.~Hoffmann, F.~Knechtli, R.~Sommer, and U.~Wolff, {\it
  {Non-perturbative renormalization of the axial current with dynamical Wilson
  fermions}},  {\em JHEP} {\bf 07} (2005) 007,
  [\href{http://xxx.lanl.gov/abs/hep-lat/0505026}{{\tt hep-lat/0505026}}].

\bibitem{Luescher1996c}
M.~Lüscher, S.~Sint, R.~Sommer, P.~Weisz, and U.~Wolff, {\it {Non-perturbative
  O(a) improvement of lattice QCD}},  {\em Nucl. Phys. B} {\bf 491} (1996) 323,
  [\href{http://xxx.lanl.gov/abs/hep-lat/9609035}{{\tt hep-lat/9609035}}].

\bibitem{Wolff2003}
U.~Wolff, {\it {Monte Carlo errors with less errors}},  {\em Comput. Phys.
  Commun.} {\bf 156} (2003) 143,
  [\href{http://xxx.lanl.gov/abs/hep-lat/0306017}{{\tt hep-lat/0306017}}].

\bibitem{Joswig2022}
F.~Joswig, S.~Kuberski, J.~T. Kuhlmann, and J.~Neuendorf, {\it {pyerrors: a
  python framework for error analysis of Monte Carlo data}},
  \href{http://xxx.lanl.gov/abs/2209.14371}{{\tt 2209.14371}}.

\bibitem{Heitger2020}
J.~Heitger and F.~Joswig, {\it {The renormalised $\mathrm{O}(a)$ improved
  vector current in three-flavour lattice QCD with Wilson quarks}},  {\em Eur.
  Phys. J. C} {\bf 81} (2021) 254,
  [\href{http://xxx.lanl.gov/abs/2010.09539}{{\tt 2010.09539}}].

\bibitem{Sint2010}
S.~Sint, {\it {The chirally rotated Schrödinger functional with Wilson
  fermions and automatic O(a) improvement}},  {\em Nucl. Phys. B} {\bf 847}
  (2010) 491, [\href{http://xxx.lanl.gov/abs/1008.4857}{{\tt 1008.4857}}].

\end{thebibliography}\endgroup

\end{document}